\documentclass[12pt]{article}
\usepackage{amssymb}
\usepackage{amsfonts}
\usepackage[all]{xy}
\usepackage{color}

\def\del{\partial}

\def\d{\delta}
\def\t{\theta^{\k\l}}
\def\T{\Theta^{KL}}

\def\del{\partial}
\def\l{\lambda}
\def\L{\Lambda}
\def\s{\sigma}
\def\E{\Sigma}
\def\sbar{\bar{\sigma}}

\def\ep{\epsilon}
\def\epbar{\bar{\epsilon}}

\def\a{\alpha}

\def\g{\gamma}
\def\G{\Gamma}
\def\k{\kappa}
\def\adot{\dot{\alpha}}
\def\bdot{\dot{\beta}}

\def\kok{\sqrt{2}}

\def\l{\lambda}
\def\p{\psi}
\def\P{\Psi}
\def\f{\phi}
\def\ft{\phi^{\dag}}

\def\pbar{\bar{\psi}}
\def\Pbar{\bar{\Psi}}
\def\lbar{\bar{\lambda}}
\def\x{\xi}

\def\xbar{\bar{\xi}}

\def\be{\begin{equation}}
\def\ee{\end{equation}}
\def\beq{\begin{eqnarray}}
\def\eeq{\end{eqnarray}}

\def\hp{\hat{\psi}}
\def\hpbar{\hat{\pbar}}

\def\hl{\hat{\l}}
\def\hlbar{\hat{\lbar}}
\def\hf{\hat{\phi}}
\def\hft{\hat{\phi^\dag}}

\begin{document}
%
\date{}
\title{\textbf{Dimensional Reduction, Seiberg--Witten Map and Supersymmetry}}
%
\author{E. Ula\c{s} Saka$^1$\thanks{E-mail:ulassaka@istanbul.edu.tr} and Kayhan \"{U}lker$^2$\thanks{E-mail:kulker@gursey.gov.tr.} }
%
\maketitle
\begin{center}
\vskip-2em
$^1$\textit{Istanbul University,\\Physics Department, TR-34134  Vezneciler--Istanbul, Turkey.}\\\vskip1em

$^2$\textit{Feza G\"{u}rsey Institute, \\P.O. Box 6,
TR--34684, \c{C}engelk\"{o}y--Istanbul, Turkey.}

\end{center}
\vskip-4em
%
%
\vskip8em
\begin{abstract}

It is argued that dimensional reduction of Seiberg–-Witten map for a gauge field induces Seiberg--Witten maps for the other noncommutative fields of a gauge invariant theory. We demonstrate this observation by dimensionally reducing the noncommutative N=1 SYM theory in 6 dimensions to obtain noncommutative N=2 SYM in 4 dimensions. We explicitly derive Seiberg--Witten maps of the component fields in 6 and 4 dimensions. Moreover, we give a general method to define the deformed supersymmetry transformations that leave the actions invariant  after performing Seiberg--Witten maps. 

\end{abstract}
\vfill
\eject
%

\section{Introduction}

Quantum field theories on noncommutative (NC) space--times recently attracted much attention  mainly due to their relation to string theory \cite{sw}, although the idea is quite old \cite{snyder}. However, this relation to string theory leads to  the interesting result that certain NC gauge theories can be mapped to commutative ones \cite{sw}. This map is commonly called as Seiberg--Witten (SW) map.

The generalization of SW--map to supersymmetric NC gauge theories was considered in Ref.\cite{fl} and \cite{ck} by using superfields in canonically deformed superspace\footnote{Here, canonically deformed superspace means $x-x$ deformation. For more general discussion of deformation of superspace including $x-\theta$ and $\theta-\theta$ deformations where $\theta$ are the Grassmann coordinates of the superspace, see for instance \cite{defo}.}. The approach of \cite{fl} is to generalize the equations that lead to SW--maps to superfields in canonically deformed superspace. Though, these equations can be solved directly they are cumbersome. Indeed, their solutions are not presented in \cite{fl}. On the other hand, the solution given in \cite{ck} is non--local and does not yield the original solution of Seiberg and Witten. 

For the component field formalism of supersymmetric theories, SW maps of the component fields of the abelian theory were given in \cite{duy} by utilizing approaches given in \cite{fl} and \cite{ck}. For NC Super Yang Mills (SYM) theory these maps were already found in \cite{wp} by using a completely  different approach. In both \cite{duy} and \cite{wp} it is assumed that the NC component fields that are superpartners of a NC gauge field functionally depend on their ordinary counterparts and on the ordinary gauge field. 

As noted in \cite{wp}, when one expands the NC action up to first order in the deformation parameter $\Theta$ after performing the SW--map, the resulting action is not invariant under classical supersymmetry transformations. This fact suggests that SUSY transformations should also be deformed when these transformations are written in terms of ordinary fields after the SW--map. Such deformations are studied in \cite{duy} and \cite{pss} for abelian NC gauge theories.  

However, note also that, by relaxing the assumption that the NC component fields are just functionals of their ordinary counterparts and gauge fields, in \cite{mik} SW--maps are studied for abelian NC supersymmetric gauge theory by solving the respective equations directly for superfields in order that the yielding action is invariant under classical SUSY transformations.  

On the other hand,  use of extra dimensions in a trivial way is a fruitful method to construct theories with larger symmetries. One of the classical example is to construct SYM theories with extended SUSY in four dimensions from higher dimensional N=1 SYM theories by using dimensional reduction \cite{bss}. Therefore, it is natural to ask if dimensional reduction of gauge theories can also shed some light on the aforementioned approaches.

In this paper, we first note that, dimensional reduction of  SW--map for a NC gauge field $\hat{A}_M$ gives directly SW--map of  NC scalar fields\footnote{SW--map of an adjoint scalar field via dimensional reduction is studied also in \cite{by}.} $\hf$ in the lower dimensions by choosing the deformation parameter in a suitable way. This observation leads to the general form of the equations  whose solutions are  SW--maps for the NC fields of a gauge invariant theory. As a direct consequence, one can generalize this result to the component fields of SYM theories. The assumption used in Ref.s \cite{duy, wp} that the NC component fields depend on their ordinary counterparts and on the ordinary gauge field, i.e. $\hf=(\f ,A)$, arises naturally by using the dimensional reduction of the original SW--map.  

After performing these SW--maps to the component fields, the resulting actions are not invariant under the (ordinary) SUSY transformations. In order to have supersymmetric actions after performing the SW--maps it is clear that NC SUSY transformations $\hat{\delta}$ should also be deformed after SW--map. Following a similar approach given in \cite{duy} for  NC supersymmetric abelian gauge theory,  we present a general method to define deformed SUSY transformations $\delta$  such that

\begin{eqnarray*}
    \xymatrix@C=4ex{          \hat{\delta} \ar[rr]_-{}^-{SW-map}
      & {} & \delta = \delta_0 + \delta_1           }
  \end{eqnarray*}    

\noindent where, $\delta_0 $ is the ordinary SUSY transformations and $\delta_1$ is the deformed part of SUSY that depends on the deformation parameter $\Theta$. These transformations are consistent with the respective SW--maps of the component fields and leave the actions invariant that are found after performing the Seiberg--Witten maps.

In order to demonstrate aforementioned ideas and methods, we study the dimensional reduction of NC N=1 SYM theory in 6 dimensional NC Minkowski space which is deformed with the help of a constant deformation parameter $\Theta$, to obtain NC N=2 SYM in 4 dimensions. Our results are presented up to first order in $\Theta$.  

The paper is organized as follows. In Section 2, we study the dimensional reduction of NC N=1  SYM theory in six dimensions to four dimensions. As expected, we show that the resulting theory is NC N=2 SYM theory in four dimensions. In Section 2, we also give the NC N=2 (on--shell) SUSY transformations of the NC component fields that leave the NC N=2 SYM action invariant in four dimensions. 

In Section 3, we study the relation between the dimensional reduction procedure and SW--maps. We construct explicitly the SW--maps for component fields of supermultiplets in 6 and 4 dimensions up to first order in the deformation parameter.  

In Section 4, we write the 6 dimensional NC N=1 SYM and 4 dimensional NC N=2 SYM actions in terms of ordinary fields by using SW--maps. We then give a general method to define deformed supersymmetry transformations that leave the actions invariant  after performing the Seiberg--Witten maps. We construct explicitly the N=1 deformed SUSY transformations in six and N=2 ones in four dimensions. 

Our conclusions are presented in Section 5. 
   
\vfill
\eject


\section{Dimensional Reduction of NC SYM Theory}

The simplest noncommutative (NC) space that is extensively studied in the literature is the deformation of  D--dimensional Minkowski or Euclidean space $\mathbb{R}^D$ :

$$[{x} ^M\, ,\, {x} ^N ]=i\Theta^{MN}$$

\noindent with the help of a real constant antisymmetric parameter $\Theta$. This NC space is characterized by Moyal $*$--product : 

$$
f(x)*g(x)=f(x)g(x)+\frac{i}{2}\Theta^{MN}\del_{M}f(x)\del_{N}g(x)+ \mathcal{O}(\Theta^2) .
$$

The action of NC N=1 SYM in six dimensions can be written by replacing the ordinary product with the Moyal $*$--product\footnote{Our conventions and some useful formula are given in Appendix A.}: 

\beq\label{S6}
\hat{S}_6=tr\int d^6x \{-\frac{1}{4}\hat{F}_{MN}\hat{F}^{MN}-\frac{i}{2}\hat{\Pbar}\G^M \hat{D}_M \hat{\P}\}
\eeq

\noindent where $\hat{F}_{MN}$ is the field strength of the NC gauge field $\hat{A}_M$ and $\hat{\Psi}$ is a Weyl spinor that belongs to the same (on--shell) supermultiplet with $\hat{A}_M$ in six dimensions.

The action (\ref{S6}) is invariant under NC SUSY Transformations;

\be\label{trans6}
\hat{\d}\hat{A}_M =-\frac{i}{2}(\hat{\Pbar}\G_M \ep -\epbar\G_M \hat{\P})
\quad,\quad
\hat{\d}\hat{\P}=\E^{MN}\ep\hat{F}_{MN} \quad,\quad
\hat{\d}\hat{\Pbar} =-\epbar\E^{MN}\hat{F}_{MN}  . 
\ee

\noindent where $\ep$ and $\epbar$ are the constant parameters of N=1 SUSY in 6 dimensions.

Dimensional reduction of NC SYM can be obtained in a similar way as it is done for the ordinary case \cite{bss}. For this purpose, we let the space--time to be decomposed as $x^M=(x^\mu,x^i)$ such that the coordinates $x^i$ are the compactified ones. We let any function $f(x)$ to be function of only uncompactified coordinates $x^\mu$ i.e. $\del_i f(x)=0$. Then, the dimensional reduction of the D-dimensional NC space can be performed by choosing the deformation parameter $\Theta$ as 

\be\label{theta}
\Theta^{MN}=\left(
\begin{array}[c]{ll}
\theta^{\mu\nu} & 0\\
0 & 0 \\
\end{array}
\right),
\ee

\noindent so that the lower dimensional space--time still has the same canonical deformation,

$$[{x} ^\mu\, ,\, {x} ^\nu ]=i\theta^{\mu\nu}\, ,\, [{x} ^i\, ,\, {x} ^j ]=0$$

\noindent This choice of the deformation parameter $\Theta$ leads trivially to the Moyal product in four dimensions:

$$
f(x)*g(x)=f(x)g(x)+\frac{i}{2}\theta^{\mu\nu}\del_{\mu}f(x)\del_{\nu}g(x)+ \mathcal{O}(\theta^2) .
$$

Note that one could also choose some other form for $\Theta$. This choice would also lead to the same Moyal $*$--product in four dimensions, since any function is considered to be independent of the compactified coordinates $x^i$. However, as it will be clear in the next Section, to be able to derive consistent SW--maps via dimensional reduction the choice (\ref{theta}) is mandatory.

After performing the compactification, the action of NC N=2 SYM in four dimensions can  be obtained as\footnote{Our definitions of lower dimensional fields via dimensional reduction are given in the appendix B.}, 

\beq\label{S4}
\hat{S}_4 &=& tr\int d^4 x (\, -\frac{1}{4}\hat{F}_{\mu\nu}\hat{F}^{\mu\nu}-i\hl\s^\mu\hat{D}_\mu \hlbar -  i\hp\s^\mu\hat{D}_\mu\hpbar- \hat{D}^\mu \hf \hat{D}_\mu\hft\nonumber
\\
&&\qquad\qquad\qquad +i\kok(\hp[\hl,\hft]_* - \hlbar[\hpbar,\hf]_*)-\frac{1}{2}[\hf,\hft]^2_*\,\,)
\eeq

\noindent The action (\ref{S4}) is invariant under the (on--shell) NC N=2 SUSY transformations that can also be obtained from (\ref{trans6}) by dimensional reduction: 

\beq\label{trans4}
\d \hat{A}_\mu &= & i\x_1\s_\mu\hlbar+i\x_2\s_\mu\hpbar+i\xbar_1\sbar_\mu\hl+i\xbar_2\sbar_\mu\hp\nonumber
\\&&\nonumber\\
\d \hl &=& \s^{\mu\nu}\x_1 \hat{F}_{\mu\nu}+i\x_1[\hf,\hft]_* -i\kok\s^\mu\xbar_2\hat{D}_\mu\hf
\nonumber\\ &&\nonumber\\
\d \hp &=& \s^{\mu\nu}\x_2 \hat{F}_{\mu\nu}+i\x_2[\hf,\hft]_* +i\kok\s^\mu\xbar_1\hat{D}_\mu\hf\nonumber
\\&&\nonumber\\
\d \hf &=& \kok\x_1\hp-\kok\x_2\hl
\eeq

\noindent where $\x_1 \, , \x_2$ are the constant parameters of N=2 supersymmetry. Note that the above action (\ref{S4}) and supersymmetry transformations (\ref{trans4}) are the same with the well known ones in the undeformed space after replacing the ordinary product with $*$--product as expected.

\section{Seiberg--Witten Map via Dimensional Reduction}

Based on the fact that one can derive both conventional and noncommutative gauge theories from the same two dimensional field theory by using different regularization procedures, Seiberg and Witten \cite{sw} showed that there exists a map from a commutative gauge field to a noncommutative one, that exhibits the equivalence between the two  theories. This map is commonly called as Seiberg--Witten map (SW--map) and arises from the requirement
that gauge invariance should be preserved in the following sense : 

\be\label{sw} 
 \hat{A}(A)+\hat{\d}_{\hat{\L}}^g\hat{A}(A)=\hat{A}(A+\d_\L ^g A)
\ee

\noindent where $\d_\L ^g$ , $\hat{\d}_{\hat{\L}}^g$ are gauge transformations with infinitesimal parameters $\L$ and $\hat{\L}$ respectively, such that $\hat{\L}=\hat{\L}(A_M , \L)$. The solutions of eq.(\ref{sw}) up to first order in $\Theta$ are found to be \cite{sw},

\beq\label{swA}
\hat{A}_M(A)&=&A_M-\frac{1}{4}\T \{A_K,\del_L A_M+F_{LM}\}+ \mathcal{O}(\Theta^2) 
\\&&\nonumber\\
\hat{\L}(A,\L)&=&\L + \frac{1}{4}\T \{\del_K\L,A_L\}+ \mathcal{O}(\Theta^2) .
\eeq

SW--map of non--commutative supersymmetric gauge theories are studied in various ways in order to get Seiberg--Witten map for the other fields that are in the same supermultiplet with a gauge field. However, dimensional reduction procedure gives directly the desired SW--maps of the component fields in a supermultiplet.

In order to get these maps, note that when one dimensionally reduce the gauge field $A_M$, the components on the compactified dimensions $A^i$ behave as scalar fields.  Therefore, the dimensional reduction of the SW--map (\ref{sw}) gives the original map 

\be \label{sw-a}
\hat{A}_\mu(A) = A_\mu -\frac{1}{4}\t \{A_\k ,\del_\l A_\mu +F_{\l\mu}\}+ \mathcal{O}(\theta^2)
\ee

\noindent and also  SW--map of the scalar fields\footnote{Here, $\f$ denotes any component $A_i$ of the gauge field on compactified coordinate  or any linear combination of $A_i$.}  $\hat{\f}$ in the lower dimensional space--time that are in the adjoint representation of the gauge group\footnote{A similar way to obtain the SW--map of the scalar fields is also studied in \cite{by}.}

\be\label{sw-f}
\hat{\f}=\f-\frac{1}{4}\t\{A_\k,(\del_\l+D_\l)\f\} + \mathcal{O}(\theta^2)
\ee 

\noindent by choosing the noncommutativity parameter to have the form given in (\ref{theta}).

Here, the choice (\ref{theta}) for the deformation parameter $\Theta$ becomes clear. Since the eq.(\ref{sw}) and its solution (\ref{swA}) is valid in any dimension \cite{sw}, the dimensional reduction of SW map (\ref{swA}) should also have the same form in lower dimension, i.e. like  (\ref{sw-a}). This is only possible if one chooses the deformation parameter $\Theta$ as  (\ref{theta}).

Since, dimensional reduction of Yang--Mills action gives Yang--Mills coupled to scalar fields, the aforementioned observation leads to the fact that a non--commutative scalar field $\hf$, that couples to gauge fields in a gauge invariant way, should be written in terms of ordinary gauge fields and ordinary scalar fields. In other words,    $\hf=\hf(A,\f)$ and just like the original case \cite{sw}, to preserve the gauge invariance of the theory $\hf$ should satisfy,

\be\label{sw-denk-f}
\hat{\f}(A, \f)+\hat{\d}_{\hat{\L}}^g\hat{\f}(A,\f)=\hat{\f}(A+\d_\L ^g A,\f+\d_\L ^g \f) 
\ee

\noindent which generates directly the SW--map of the scalar fields (\ref{sw-f}). 

One can deduce that a similar argument should also hold for any field that couples to gauge fields in a gauge invariant theory. Therefore, for instance for the NC Weyl spinors $\hat{\Psi}$ in (\ref{S6}) one can write a  similar condition like Eq.(\ref{sw-denk-f}) that gives SW--map of $\hat{\Psi}$ :  

\be\label{swP}
\hat{\P}=\P-\frac{1}{4}\T \{A_K,(\del_L+ D_L)\P \}+ \mathcal{O}(\Theta^2)
\ee

Clearly, SW--maps of the Weyl spinors $\hat{\p}$ and $\hat{\l}$ of NC N=2 SYM in 4 dimensions have the same form :

\be\label{sw-p}
\hat{\p}=\p-\frac{1}{4}\t\{A_\k,(\del_\l+D_\l)\p\}+\mathcal{O}(\theta^2) \quad,\quad\hat{\l}=\l-\frac{1}{4}\t\{A_\k,(\del_\l+D_\l)\l\}+ \mathcal{O}(\theta^2) 
\ee

These SW--maps (\ref{swP}) and (\ref{sw-p}) are also consistent with each other in the sense that (\ref{sw-p}) can be obtained from the dimensional reduction of (\ref{swP}) when the deformation parameter is chosen as (\ref{theta}). 

Obviously, the maps (\ref{swA},\ref{swP}) and (\ref{sw-a},\ref{sw-f},\ref{sw-p}) gives the desired SW--maps for the component fields of N=1 NCSYM in 6 dimensions and N=2 NCSYM in 4 dimensions respectively. Moreover, the above derived SW--maps are in agreement with the ones that are found for the components of four dimensional N=1  supersymmetric gauge theories in \cite{duy} for U(1) case and in \cite{wp} for nonabelian case. However, our method, which is considerably simpler, and the methods studied in \cite{duy} and \cite{wp} to find these maps are completely different from each other.

On the other hand, the assumption that the NC component fields in a supersymmetric gauge theory depend on gauge field and their ordinary counterparts \cite{duy,wp}, arises here naturally as a direct consequence of the dimensional reduction of SW--map of the gauge field $A_M$. Indeed, one can  generalize the above observation for any NC field that couples to a gauge field whether the theory is supersymmetric or not.

\section{Deformed SUSY Transformations }

After obtaining the SW--maps (\ref{swA}) and (\ref{swP}) of the component fields of NC SYM theory in six dimensions one can write the action (\ref{S6}) in terms of ordinary component fields $(A_M ,\Psi , \bar{\Psi})$ up to order $\Theta$ as

\beq\label{Sdef6}
S_6 &=& tr\int d^4x (-\frac{1}{4}F_{MN}F^{MN}-\frac{i}{2}\Pbar\G^M D_M \P -\frac{1}{4}\T (F^{MN}\{F_{MK},F_{NL}\}-\frac{1}{4} F_{KL}\{F_{MN},F^{MN}\}\nonumber
\\
&&\qquad \qquad\quad -\frac{i}{2} \Pbar\G^M(\{F_{KL},D_M\P\}+2\{F_{MK},D_L\P\}) )
\eeq

N=2 SYM action in 4 dimensions can be derived in two different ways:  either by dimensionally reducing the action (\ref{Sdef6}) or by applying the SW--maps (\ref{sw-a},\ref{sw-f},\ref{sw-p}) to the action (\ref{S4}). One can show that both ways of obtaining the action give exactly the same result : 

\beq\label{Sdef4}
S_4 &=& tr \int d^4 x (\,-\frac{1}{4}F_{\mu\nu}F^{\mu\nu}-i\l\s^\mu D_\mu\lbar-i\p\s^\mu D_\mu\pbar- D^\mu\f D_\mu\ft
\nonumber\\&&\qquad\qquad\qquad \qquad\qquad
+i\kok(\p[\l,\ft] -\lbar[\pbar,\f] )-\frac{1}{2}[\f,\ft]^2 \,\,) \nonumber\\
&+&   tr\int d^4 x\,\t\,  (\, -\frac{1}{4}F^{\mu\nu} \{F_{\mu \k},F_{\nu \l}\} + \frac{1}{16}F_{\k\l}\{F_{\mu\nu},F^{\mu\nu}\}\nonumber
\\
&&\qquad\quad + \frac{i}{4}\l\s^\mu(\{F_{\k\l},D_\mu \lbar\}+2\{F_{\mu \k},D_\l \lbar\})
 + \frac{i}{4}\p\s^\mu(\{F_{\k\l},D_\mu \pbar\}+2\{F_{\mu\k},D_\l \pbar\})\nonumber
\\
&&\qquad\quad -\frac{1}{2}(D^\mu\ft\{F_{k\mu},D_\l\f\}+D^\mu\f\{F_{\k\mu},D_\l\ft\})
+\frac{1}{4}F_{\k\l}\{D_\mu\ft,D^\mu\f\} -\frac{i\kok}{4}\{F_{\k\l},\p\}[\l,\ft]\nonumber
\\
&&\qquad\quad -\frac{\kok}{2}\p\{D_\k\l,D_\l\ft\} +\frac{i\kok}{4}\{F_{\k\l},\lbar\}[\pbar,\f]+\frac{\kok}{2}\lbar\{D_\k\pbar,D_\l\f\}\nonumber
\\
&&\qquad\quad +\frac{1}{8}[\ft,\f]\{F_{\k\l},[\ft,\f]\}-\frac{i}{2}[\ft,\f]\{D_\k\ft,D_\l\f\}\,\,)
\eeq

It is clear that, neither the action (\ref{Sdef6}) nor (\ref{Sdef4}) is invariant under the respective classical SUSY transformations\footnote{The ordinary SUSY transformations can be read easily from the transformations  (\ref{trans6}) and (\ref{trans4}) by replacing the NC fields with the ordinary ones and *--product with the ordinary product. }. This was first noted in \cite{wp}.  To over come this circumstance one can deform the NC SUSY generators $\hat{\d}$. Such deformations are studied in \cite{duy,pss} for abelian NC supersymmetric gauge theories. 

However, following \cite{duy}, one can attain a general method to construct deformed SUSY transformations that keep the deformed supersymmetric gauge theory actions invariant. To achieve this goal, we let the generator of the SUSY transformation $\hat{\delta}$  to be :

\begin{eqnarray*}
    \xymatrix@C=4ex{      \hat{\d} \ar[rr]_-{}^-{SW-map}
      & {} & \delta = \delta_0 + \delta_1
    }
  \end{eqnarray*} 

\noindent where $\delta_0 $ is the ordinary SUSY transformations and $\delta_1$ is the deformed part at the order of $\Theta$. The invariance of a NC action $\hat{S}$ under NC SUSY transformations $\hat{\d}$ is then mapped to the invariance of the action $S$ under the new deformed SUSY transformations $\d$ after implementing SW maps :

\begin{eqnarray*}
    \xymatrix@C=4ex{          \hat{\delta}\hat{S}(\hat{\Phi}) =0 \ar[rr]_-{}^-{SW-map}
      & {} & \delta S(\Phi;\Theta) =0           }
  \end{eqnarray*} 

\noindent where $\hat{\Phi}$ and $\Phi$ denotes collectively all the NC component fields and their ordinary counterparts respectively including the gauge field.

In order to construct consistent SUSY transformations with  SW--maps of the component fields, let us denote the NC supersymmetry transformations as

$$\hat{\delta} \hat{\Phi} = \hat{X}$$

\noindent where $\hat{X}$ denotes  SUSY transformation of a NC field $\hat{\Phi}$. After performing the SW--map on both sides of the transformation one can write up to first order in $\Theta$,

$$\hat{\delta} \hat{\Phi} = \delta_0\Phi + \delta_0\Phi^{(1)} + \delta_1 \Phi + \mathcal{O}(\Theta^2) = X +X^{(1)} + \mathcal{O}(\Theta^2) $$ 

\noindent which leads to 

$$\delta_0\Phi = X \, , \, \delta_0\Phi^{(1)} + \delta_1 \Phi = X^{(1)}$$

\noindent where $\hat{\Phi} = \Phi + \Phi^{(1)} +\cdots \, ,\, \hat{X}=X+X^{(1)}+\cdots$ and $\Phi^{(1)}\, ,\,X^{(1)}$ denote the first order terms in $\Theta$ after SW--map. It is clear that $\delta_0\Phi = X$ are ordinary SUSY transformations. On the other hand, since $\Phi^{(1)}$ is a polynomial of (ordinary) component fields $\Phi$ and their derivatives, and  since $\delta_0$ transformation of $\Phi^{(1)}$ is already known, one can read the action of the generator $\delta_1$ on the fields $\Phi$ as
$\delta_1 \Phi =  X^{(1)} - \delta_0\Phi^{(1)}$. Deformed SUSY transformations $\d$ of the (ordinary) component fields can then be written as 

\be
\d \Phi = X + X^{(1)} - \delta_0\Phi^{(1)} .
\ee

Deformed SUSY transformations of six dimensional NC N=1 SYM  can be derived by following the above given steps. The resulting transformations are then found to be,

\beq\label{transdef6}
&\d A_M& = -\frac{i}{2}(\Pbar\G_M \ep -\epbar\G_M \P) +\frac{i}{8}\T(\{A_K,D_M (\Pbar \G_L\ep-\epbar \G_L\P)\}\nonumber\\
&&\qquad\quad +\{\Pbar \G_L\ep-\epbar \G_L\P,\del_KA_M+F_{KM}\})
\nonumber\\&&\nonumber\\
&\d\P& = \E^{MN}\ep F_{MN} +\frac{1}{2}\T ( \E^{MN} \ep \{F_{MK},F_{NL}\} \nonumber\\
&&\qquad\quad +\frac{i}{4}\{\Pbar \G_L\ep-\epbar \G_L\P,(D_K+\del_K)\P\}
-\frac{1}{4} \{A_K,[\Pbar \G_L\ep-\epbar \G_L\P,\P]\})
\nonumber\\&&\nonumber\\
&\d\Pbar& =-\epbar\E^{MN}F_{MN}  -\frac{1}{2}\T (\epbar\E^{MN}\{F_{MK},F_{NL}\}\nonumber\\
&&\qquad\quad-\frac{i}{4}\{\Pbar \G_L\ep-\epbar \G_L\P,(D_K+\del_K)\Pbar\}
+\frac{1}{4}\{A_K,[\Pbar \G_L\ep-\epbar \G_L\P,\Pbar]\}).
\eeq

The deformed SUSY transformations $\delta$ that are constructed by using the aforementioned procedure, automatically guarantees that the action $S(\Phi;\Theta)$ is invariant under $\delta$. This is due to the fact that the transformations $\delta$ are derived directly from the NC SUSY transformations $\hat{\delta}$ that leaves a NC action invariant, $\hat{\delta}\hat{S}(\hat{\Phi})=0$. 

Indeed, one can check explicitly that the above given deformed SUSY transformations in six dimensions (\ref{transdef6})  leave the deformed  N=1 SYM action (\ref{Sdef6}) invariant. 

Note also that, the above transformations (\ref{transdef6}) are also consistent with the SW maps (\ref{swA}) and (\ref{swP}) by construction. This consistency can also be checked  explicitly, by applying directly SW maps on both sides of the NC transformations (\ref{trans6}).

On the other hand, one can follow two equivalent ways to obtain the deformed SUSY transformations  of N=2 NC SYM that leaves (\ref{Sdef4}) invariant in four dimensions:  either by dimensionally reducing the transformations (\ref{transdef6}) or from the NC N=2 SUSY transformations (\ref{trans4}) by using the aforementioned method. Both approaches give the same result : 

\beq\label{transdef4}
&\d A_\mu &= i\x_1\s_\mu\lbar+i\x_2\s_\mu\pbar+i\xbar_1\sbar_\mu\l+i\xbar_2\sbar_\mu\p \nonumber\\
&&\quad+ \frac{i}{4}\t( \{\x_1\s_\k\lbar+\x_2\s_\k\pbar+\xbar_1\sbar_\k\l+\xbar_2\sbar_\k\p,\del_\l A_\mu+F_{\l\mu}\} \nonumber \\
&&\qquad \quad
+ \{A_\l,D_\mu(\x_1\s_\k\lbar+\x_2\s_\k\pbar+\xbar_1\sbar_\k\l+\xbar_2\sbar_\k\p)\}) \nonumber
\\ &&\nonumber\\
&\d \l &=\s^{\mu\nu}\x_1 F_{\mu\nu}+i\x_1[\f,\ft]-i\kok\s^\mu\xbar_2 D_\mu\f
\nonumber
\\
&&\quad + \frac{1}{2}\t(\x_1\s^{\mu\nu}\{F_{\mu \k},F_{\nu \l}\}+i \kok\xbar_2 \{F_{\mu \k},D_\l\f\}+ \x_1 \{D_\k\ft,D_\l\f\} \nonumber
\\
&&\qquad+\frac{i}{2} \{\x_1\s_\k\lbar+\x_2\s_\k\pbar + \xbar_1\sbar_\k\l+\xbar_2\sbar_\k\p,(\del_\l+D_\l)\l\}\nonumber
\\
&&\qquad\quad+\frac{1}{2}\{A_\k,[\x_1\s_\l\lbar+\x_2\s_\l\pbar+\xbar_1\sbar_\l\l+\xbar_2\sbar_\l\p,\l]\})\nonumber
\\&&\nonumber\\
&\d \p &=\s^{\mu\nu}\x_2 F_{\mu\nu}+i\x_2[\f,\ft]+i\kok\s^\mu\xbar_1 D_\mu\hf\nonumber
\\
&&\quad -\frac{1}{2}\t(\x_2\s^{\mu\nu}\{F_{\mu \k},F_{\nu \l}\}-i\kok \xbar_1\{F_{\mu \k},D_\l\f\} + \x_2\{D_\k\f,D_\l\ft\}\nonumber
\\
&&\qquad -\frac{i}{2}\{\x_1\s_\k\lbar+\x_2\s_\k\pbar+\xbar_1\sbar_\k\l+\xbar_2\sbar_\k\p,(\del_\l+D_\l)\p\}\nonumber
\\
&&\qquad\quad -\frac{1}{2}\{A_\k,[\x_1\s_\l\lbar+\x_2\s_\l\pbar+\xbar_1\sbar_\l\l+\xbar_2\sbar_\l\p,\p]\})\nonumber
\\&&\nonumber\\
&\d \f &=\kok\x_1\p-\kok\x_2\l \nonumber\\
&&\quad + \frac{i}{4}\t(\{\x_1\s_\k\lbar+\x_2\s_\k\pbar+\xbar\sbar_\k\l+\xbar_2\sbar_\k\p,(\del_\l+D_\l)\f\}\nonumber
\\
&&\qquad\quad -i\{A_\k,[\x_1\s_\l\lbar+\x_2\s_\l\pbar + \xbar_1\sbar_\l\l +\xbar_2\sbar_\l\p ,\f]\}).
\eeq

\section{Conclusion}

We argued that the use of extra dimensions in a trivial way for non--commutative gauge theories leads to the  equations

\beq
\hat{A}(A)+\hat{\d}_{\hat{\L}}^g\hat{A}(A)&=&\hat{A}(A+\d_\L ^g A)\\\label{swO}
\hat{\Omega}(A, \Omega)+\hat{\d}_{\hat{\L}}^g\hat{\Omega}(A,\Omega) &=& \hat{\Omega}(A+\d_\L ^g A,\Omega+\d_\L ^g \Omega) 
\eeq

\noindent where the first equation is the original one for gauge fields derived in \cite{sw} and the second one (\ref{swO}) is for any NC field $\hat{\Omega}$ which is not the gauge field in a gauge invariant theory. Note that, in order to keep the form of SW map of the gauge field $A$ same before and after performing the dimensional reduction, the deformation parameter has to be chosen as in (\ref{theta}). 

As a direct consequence of the dimensional reduction of the original SW--map, one can prove that the non-commutative fields $\hat{\Omega}$ depend on their ordinary counterparts $\Omega$ and also on the gauge field $A$. The solutions of these equations are  SW--maps of the respective fields. It is clear that this result lets one to write the SW maps of the component fields in NC SYM theories. We gave these SW--maps for the  component fields of NC N=1 SYM in 6 and the ones of NC N=2 SYM in 4 dimensions.

It is also worth mentioning here that one can perform dimensional reduction from a  NC space to another before or after performing SW--map. We pointed out this observation for the dimensional reduction of the NC N=1 SYM in 6 dimensions that can be expressed by the following commuting diagram:

\begin{eqnarray*}
    \xymatrix@C=10ex{      \hat{S}_6 \ar[d]_-{Dim.Red.}
      \ar[rr]_-{}^-{SW-map}
      & {} & S_6
      \ar[d]^-{Dim.Red.}_-{} \\
      \hat{S}_4 \ar[rr]_-{SW-map} & {} &
      S_4              }
\end{eqnarray*}

On the other hand, when one considers NC SYM on the level of the ordinary component fields,  to keep the SUSY invariance of NC SYM actions after SW--map, one must also deform  the SUSY transformations after SW--map :

\begin{eqnarray*}
    \xymatrix@C=4ex{      \hat{\delta} \ar[rr]_-{}^-{SW-map}
      & {} & \delta = \delta_0 + \delta_1     }
  \end{eqnarray*} 

\noindent We gave a general method to construct such deformed SUSY transformations once SW--maps of the component fields are known. These transformations are consistent with  SW maps and leave the NC action invariant after performing  SW--map by construction. We gave two such examples, namely the deformed (on-shell) SUSY transformations of NC N=1 SYM in 6 dimensions and NC N=2 SYM in 4 dimensions. We have explicitly derived the  deformed SUSY transformations for these two cases that leave the respective actions invariant after SW--map.

Finally, it is also worth to note that, the method given in Section 4 to construct deformations of SUSY transformations in NC spacetimes is quite general.  We expect that one can generalize this procedure to other NC supersymmetric models, such as NC supergravity\footnote{Work in progress.}.

\begin{center}
{\bf {Acknowledgments:}}
\end{center}
We thank \"Omer F. Day\i{} and Cemsinan Deliduman for various enlightening discussions.

\appendix

\section{Conventions : }

The covariant derivative and the field strength are given in the non--commutative space as

\beq
\hat{D}_M \hat{\P}&=&\del_M \hat{\P}-i[\hat{A}_M,\hat{\P}]_* ,\nonumber\\
\hat{F}_{MN}&=&\del_M \hat{A}_N-\del_N \hat{A}_M-i[\hat{A}_M,\hat{A}_N]_*
\eeq

The Moyal $*$--Product is defined as;

\be
f(x)*g(x)=f(x)g(x)+\frac{i}{2}\Theta^{MN} \del_{M}f(x) \del_{N} g(x) + \mathcal{O}(\Theta^2)
\ee

\noindent whereas $[A,B]_*=A*B-B*A$ is the $*$--commutator.

Our SUSY conventions in six dimensions are similar with that of \cite{dv}.  The gamma matrices in 6 dimensions satisfy,

$$\{\Gamma_M , \Gamma_N \}= -2 \eta_{MN}$$

\noindent where $
\eta_{MN}=(-1,1,1,1,1,1).$

We often use the identities,

\beq 
\Gamma_M \E_{RS} &=& -\frac{1}{2}(\eta_{MR} \Gamma_{S}-\eta_{MS} \Gamma_{R}-\frac{i}{6} \epsilon_{MRSABC} \Gamma^{7}\Gamma^{A}\Gamma^{B}\Gamma^{C})\nonumber
\\ 
\E_{RS}\Gamma_{M} &=&  -\frac{1}{2}(-\eta_{MR} \Gamma_{S}+\eta_{MS} \Gamma_{R}+ \frac{i}{6} \epsilon_{MRSABC} \Gamma^{7}\Gamma^{A}\Gamma^{B}\Gamma^{C})\nonumber
\eeq

\noindent throughout the calculations where $\E_{MN}=\frac{1}{4}[\G_M,\G_N]$ and $\epsilon$ is the totally anti-symmetric tensor.

In four-dimensions, we use Wess-Bagger conventions \cite{wessbagger}, i.e.
      
$$\l^\a =\epsilon^{\a\beta}\l_\beta \quad,\quad \l_\a =\epsilon_{\a\beta}\l^\beta \quad,\quad \l\psi = \l^\a\psi_\a = \psi\l$$
$$\lbar^{\adot} =\epsilon^{\adot\bdot}\lbar_{\bdot} \quad,\quad 
\lbar_{\adot} =\epsilon_{\adot\bdot}\lbar^{\bdot}
\quad,\quad  
\lbar \pbar = \lbar_{\adot} \pbar^{\adot} =\pbar\lbar$$
$$\epsilon^{12}=-\epsilon^{21}=-\epsilon_{12}=\epsilon_{21}=1$$\\
$$
\sigma^0 = \left( \begin{array}{cc} -1 & 0 \\ 0 & -1 \end{array} \right)
,
\sigma^1 = \left( \begin{array}{cc} 0 & 1 \\ 1 & 0 \end{array} \right)
,
\sigma^2 = \left( \begin{array}{cc} 0 & -i \\ i & 0 \end{array} \right)
,
\sigma^3 = \left( \begin{array}{cc} 1 & 0 \\ 0 & -1 \end{array} \right).
$$

$$
\sigma^\mu_{\: \alpha \dot \alpha} = 
\epsilon_{\alpha \beta} \epsilon_{\dot \alpha \dot \beta}
{\bar \sigma}^{\mu \: {\dot \beta} \beta}
\quad,\quad
{\bar \sigma}^{\mu \: {\dot \alpha} \alpha} =
\epsilon^{\dot \alpha \dot \beta} \epsilon^{\alpha \beta}
\sigma^\mu_{\: \beta \dot \beta} 
$$
$$
\sigma^{\mu \nu \: \beta}_{\: \alpha}=
\frac{1}{4} (\sigma_{\alpha \adot}^\mu \sbar^{{\adot} \beta \: \nu}
- \sigma_{\alpha \adot}^\nu \sbar^{{\adot} \beta \: \mu}) 
\quad,\quad
\sbar^{\mu \nu \: \adot}_{\: \quad \dot \beta}=
\frac{1}{4} (\sbar^{{\adot} \alpha \: \mu} \sigma^\nu_{\alpha \dot\beta}
- \sbar^{{\adot} \alpha \: \nu} \sigma^\mu_{\alpha \dot\beta})
$$

\section{Dimensional Reduction : }

To perform the dimensional reduction we use the following parametrization of the $\Gamma$ matrices,  

\be
\G^\mu=I\otimes\g^{\mu},\; \G^{4}=\s^{1}\otimes\g^{5},\;\G^{5}=\s^{2}\otimes\g^{5}, \;
\G^{7}=\s^{3}\otimes\g^{5}
\ee

\noindent where $\s$'s are Pauli matrices and $\gamma^5=\gamma^0\gamma^1\gamma^2\gamma^3$. In this parametrization,  the Weyl spinor in six dimensions can be written in terms of four dimensional Dirac spinors:

\begin{displaymath}
\P=\left(
\begin{array}{l}
(\frac{1+i\g_{5}}{2})\chi \\
(\frac{1-i\g_{5}}{2})\chi
\end{array}
\right)\quad , \quad
\Pbar=\left(
\begin{array}{ll}
 \bar{\chi}(\frac{1-i\g_{5}}{2})& \bar{\chi}(\frac{1+i\g_{5}}{2})
\end{array}
\right)
\end{displaymath}

The Weyl spinors of the N=2 supermultiplet are then obtained as;

\begin{displaymath}
\chi=\kok \left(
\begin{array}{l}
\l_{\a} \\
\pbar^{\adot}
\end{array}
\right)
\quad , \quad
\bar{\chi}=\kok \left(
\begin{array}{ll}
 \p^{\a}& \lbar_{\adot}
\end{array}
\right)
\end{displaymath}

\noindent whereas the scalar field and its hermitian conjugate are defined in our notation as ;

\be
\f =-\frac{1}{\kok}(A_4-iA_5) ,\;\;  \ft=-\frac{1}{\kok}(A_4+iA_5). \nonumber
\ee


\end{document}